\documentstyle[prb,aps,floats,psfig]{revtex}

\long\def\singlecol#1{
\twocolumn[\hsize\textwidth\columnwidth\hsize\csname @twocolumnfalse\endcsname
              #1]}

\ifpreprintsty\long\def\singlecol#1{#1}\fi

\def\figref#1{{\protect{\ref{#1}}}}

\newcommand{\pcite}[2]{#2\cite{#1}}
\ifpreprintsty\renewcommand{\pcite}[2]{\cite{#1}#2}\fi

\long\def\beginfigeps#1#2{\begin{figure}[htb] 
   \centerline{\psfig{file=#1,width=8.4cm}}
    \protect{#2}
 \end{figure}}

\def\showfigures{}

\ifpreprintsty
      \long\def\beginfigeps#1#2{}
      
      \def\showfigures{
           \long\def\beginfigeps##1##2{\begin{figure}
   \centerline{\psfig{file=##1,width=14cm}} ##2 \end{figure}}
\allfigures}
\fi

\def\ch{\rho}

\def\kr{K_{\ch}}

\def\eps{\varepsilon}

\def\vec#1{\hbox{\mathversion{bold}${#1}$}}

\def\om{\omega}

\def\SPSF{\ single-particle spectral function\ }
\def\QMC{\ Quantum-Monte-Carlo\ }
\def\LM{\ Luttinger model\ }

\long\def\taglia#1{}

\newcommand{\cmag}{\gtrsim}

\def\beq{\begin{equation}}
\def\eeq{\end{equation}}
\def\beqn{\begin{eqnarray}}
\def\eeqn{\end{eqnarray}}

\def\eqref#1{ 
 (\ref{#1})}

\def\chirho{\beginfigeps{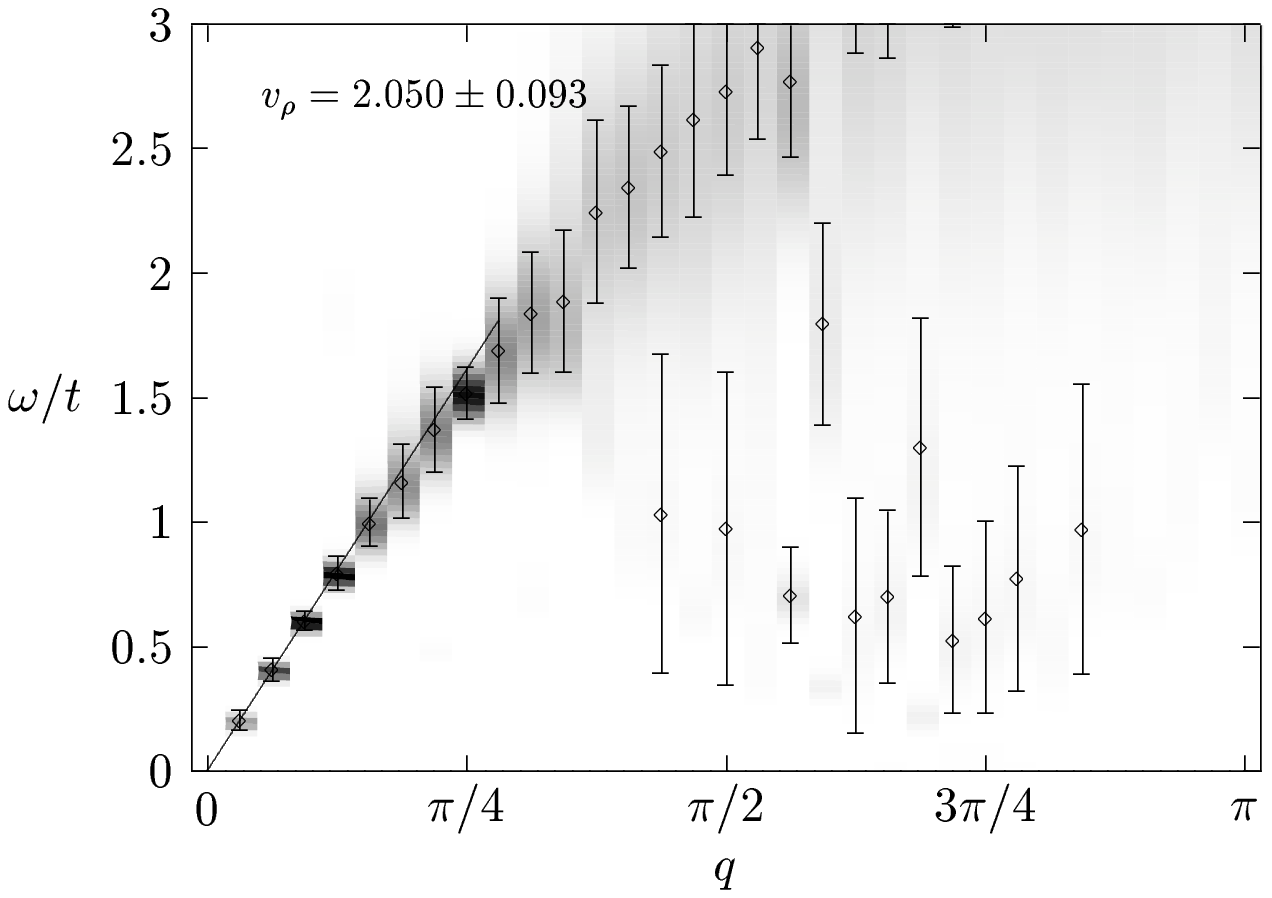}{
\caption{Density plot of the charge susceptibility $\chi_\rho
  (q,\omega)$ as obtained by the
analytic continuation of the \QMC charge-charge correlation function
with the Maximum-Entropy method. The grayscale
corresponds to the value of $\chi_\rho (q,\omega)$ 
(darker regions correspond to larger values of $\chi_\rho (q,\omega)$)
and the dots with
errorbars 
show the
 peak position with their uncertainty. 
The linear fit (straight line) for small q yields the
charge velocity $v_\rho$ as indicated in the upper left corner.}
\label{chirho}}}

\def\chisigma{\beginfigeps{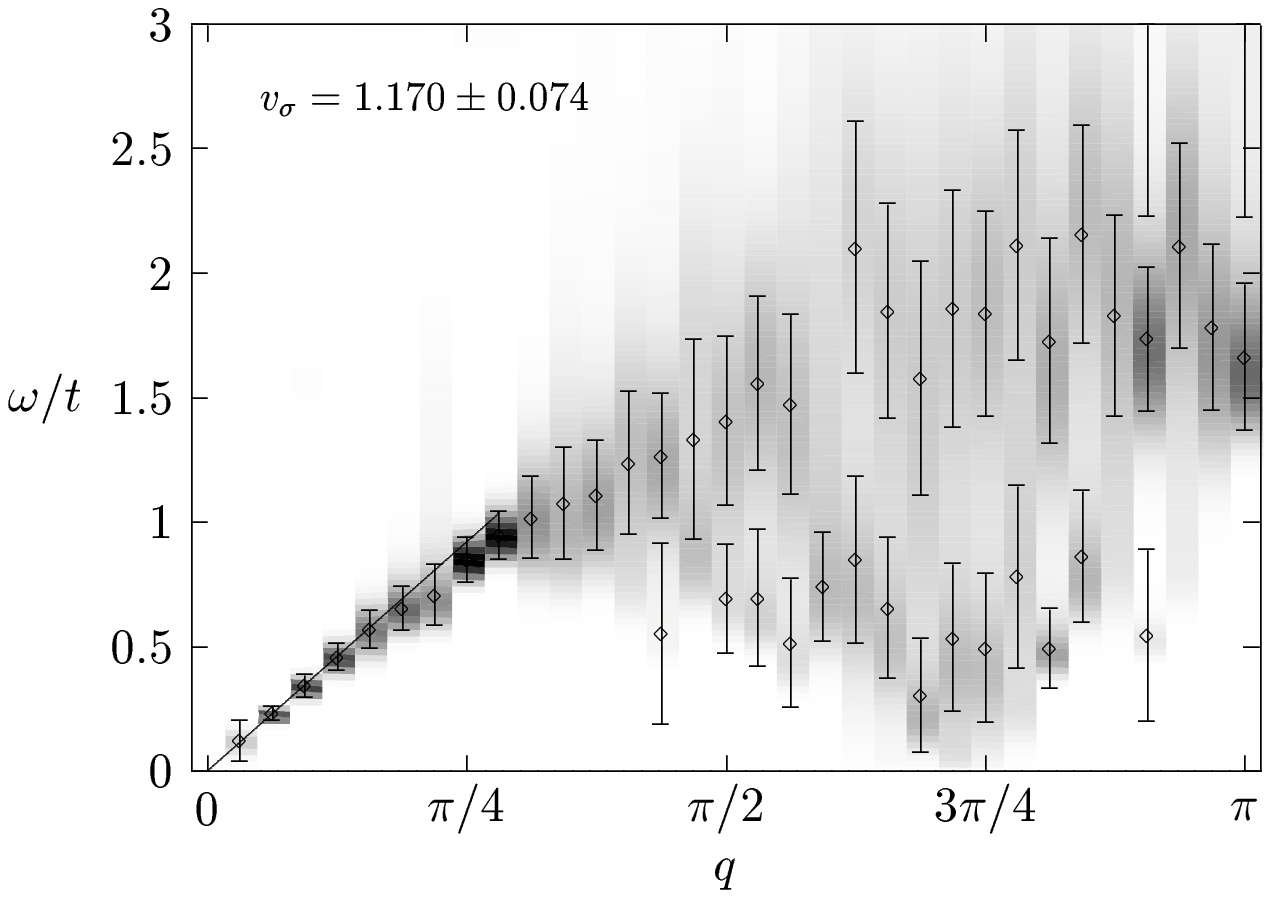}{
\caption{Density plot of the spin susceptibility $\chi_\sigma
  (q,\omega)$ 
with the same conventions as Fig. \figref{chirho}
 The linear fit (straight line) 
 yields the
spin velocity $v_\sigma$.}
\label{chisigma}}}

\def\ako{\beginfigeps{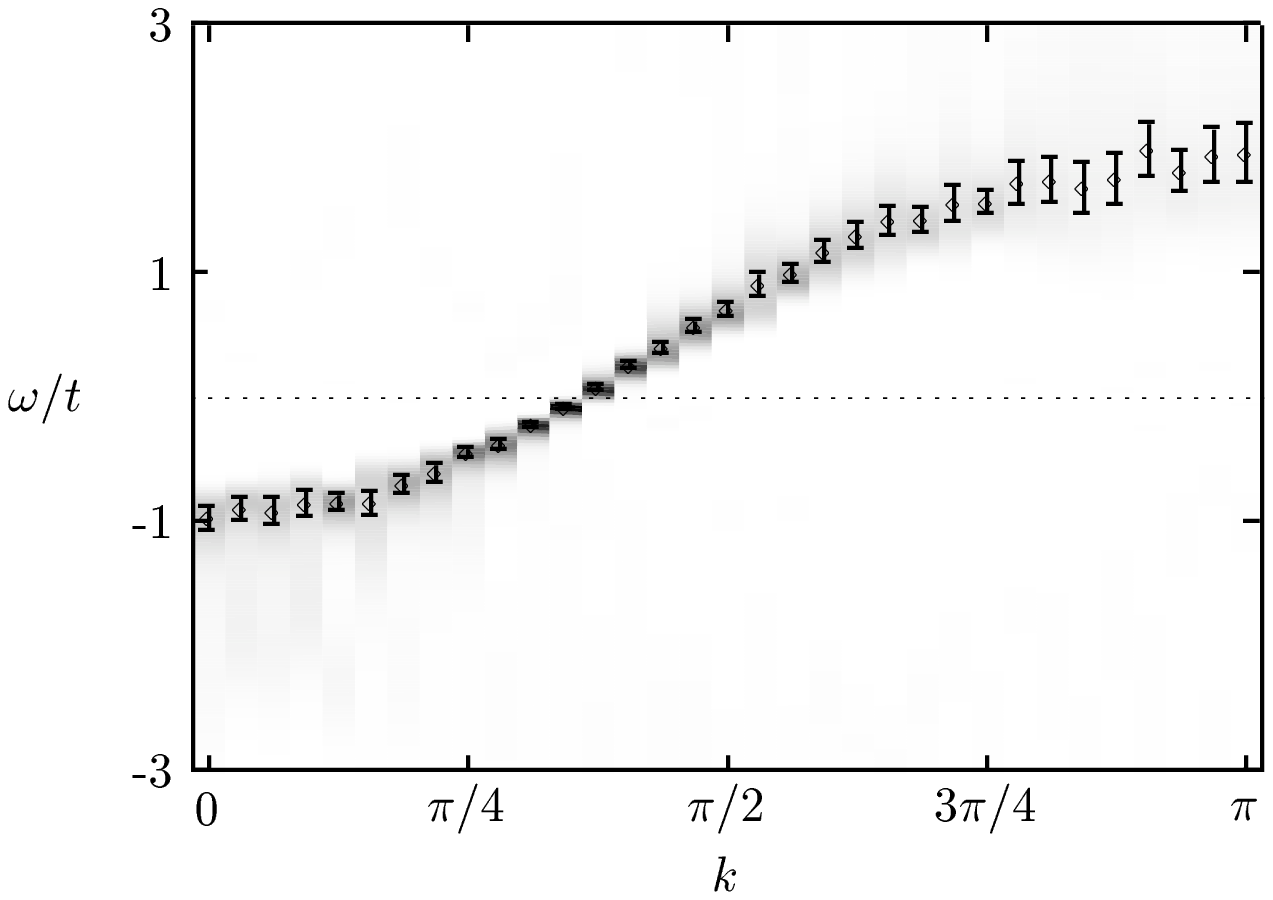}{
\caption{Density plot of the single-particle photoemission spectrum $A(k,\omega)$
with the same conventions as Fig. \figref{chirho}.
 It is seen that the dispersion around the Fermi energy
(dotted line) is linear over a broad momentum range thus justifying our {\it Luttinger-liquid}
ansatz for the single-particle Green's function.}
\label{ako}}}

\def\aksone{\beginfigeps{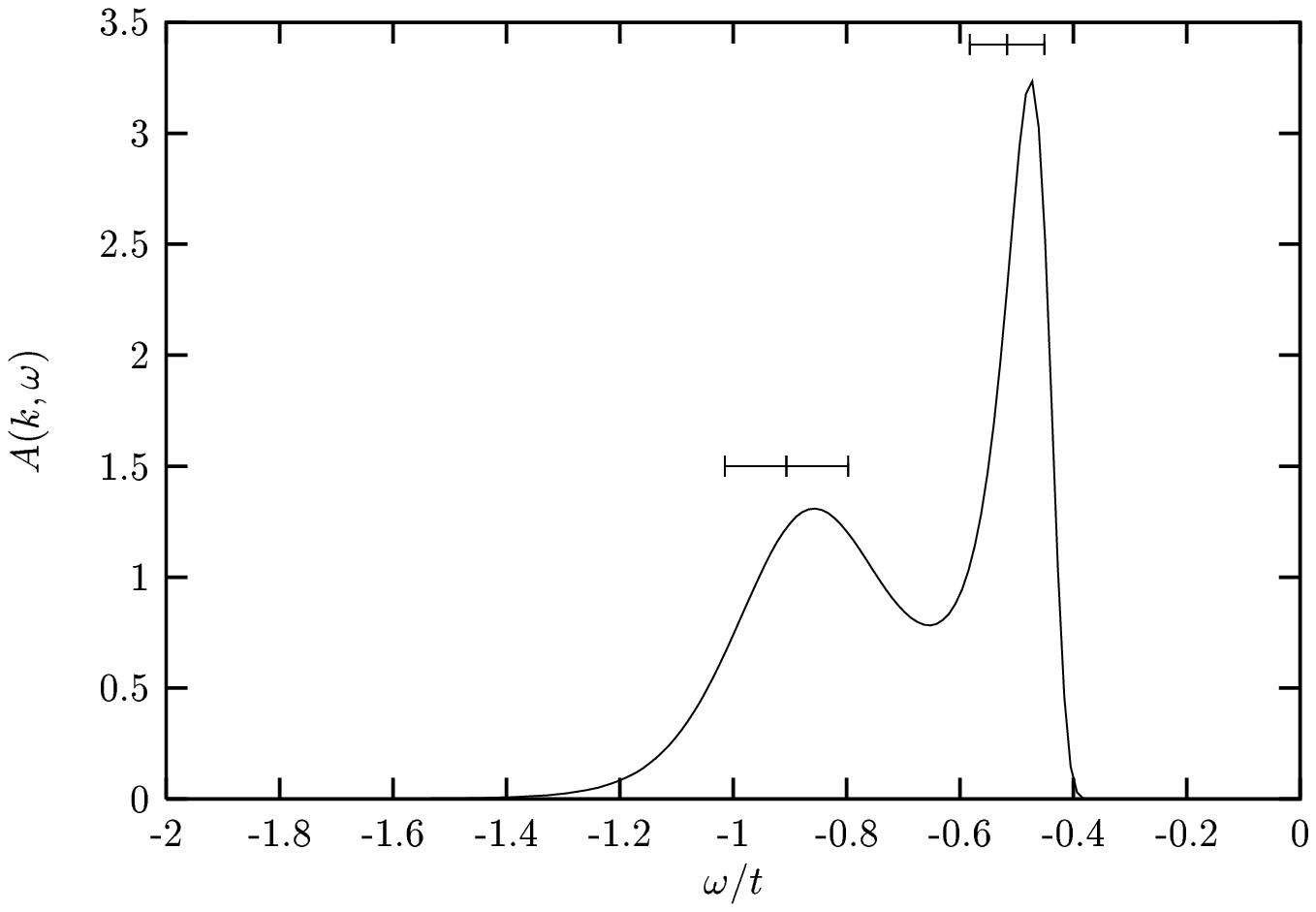}{
\caption{Single-particle photoemission spectrum $A(k,\omega)$ (in arbitrary units)
 for $k-k_F=-4.5 \,
\frac{\pi}{32}$. 
The dots with horizontal errorbars indicate the position of 
spin- and charge-excitations calculated by 
$\omega_{\rho/\sigma} = (k-k_F) \, v_{\rho/\sigma}$ with 
$v_{\rho/\sigma}$ obtained from Figs. \figref{chirho} and
\figref{chisigma}. 
For this $k$-point close to the
Fermi momentum the Maximum-Entropy method is able to resolve two separate peaks in the
spectral function which can be identified as the spinon and holon
excitation, respectively.
}
\label{aks1}}}

\def\figlog{\beginfigeps{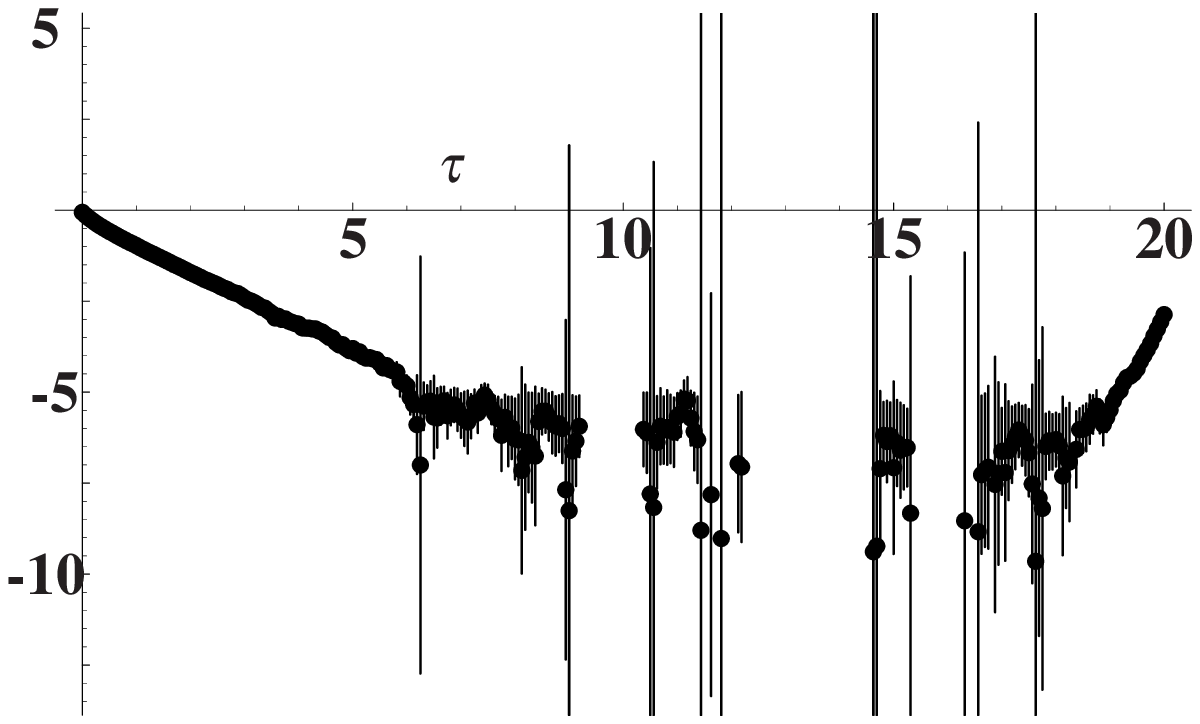}{
\caption{
Logarithmic plot of 
 the imaginary-time Green's function ${\cal G}(k,\tau)$ vs 
  $\tau$ with $k=k_F + \pi 9/64$, as obtained from \QMC data. 
}
\label{figlog}
}}

\def\figfit{\beginfigeps{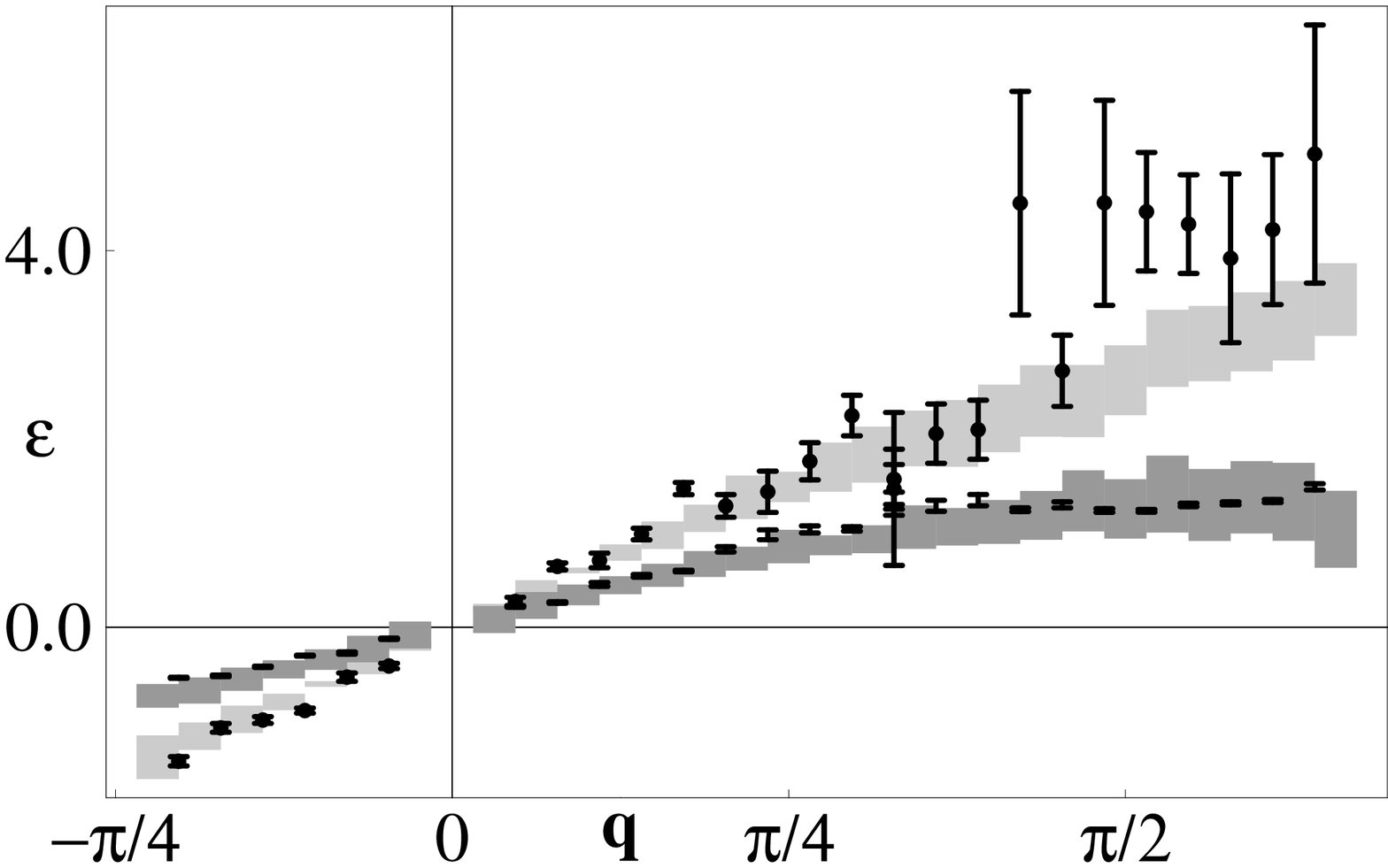}{
\caption{
Spin and charge dispersions
$\eps_1$ and  $\eps_2$ vs $q=k-k_F$
 (errorbars without and
with central dot, respectively)
as obtained 
from the $\chi^2$ 
fit of the \QMC data for the imaginary-time Green's function  
with the Luttinger liquid
Green's function [\eqref{glmk} with \eqref{glmx}]. 
The Luttinger liquid Green's function is taken at 
zero
temperature and with correlation exponent $\kr=1$.
The fit is carried out for the data in the time interval $1.0 \le \tau \le 5.0$.
For comparison, we also show the dispersions obtained 
from the 
peak positions (with corresponding uncertainty) of 
the spin (dark gray)
and of the charge  (light gray) dispersions
 (Cfr. Figs. \figref{chisigma} and \figref{chirho}).
}
\label{figfit}
}}

\def\figten{\beginfigeps{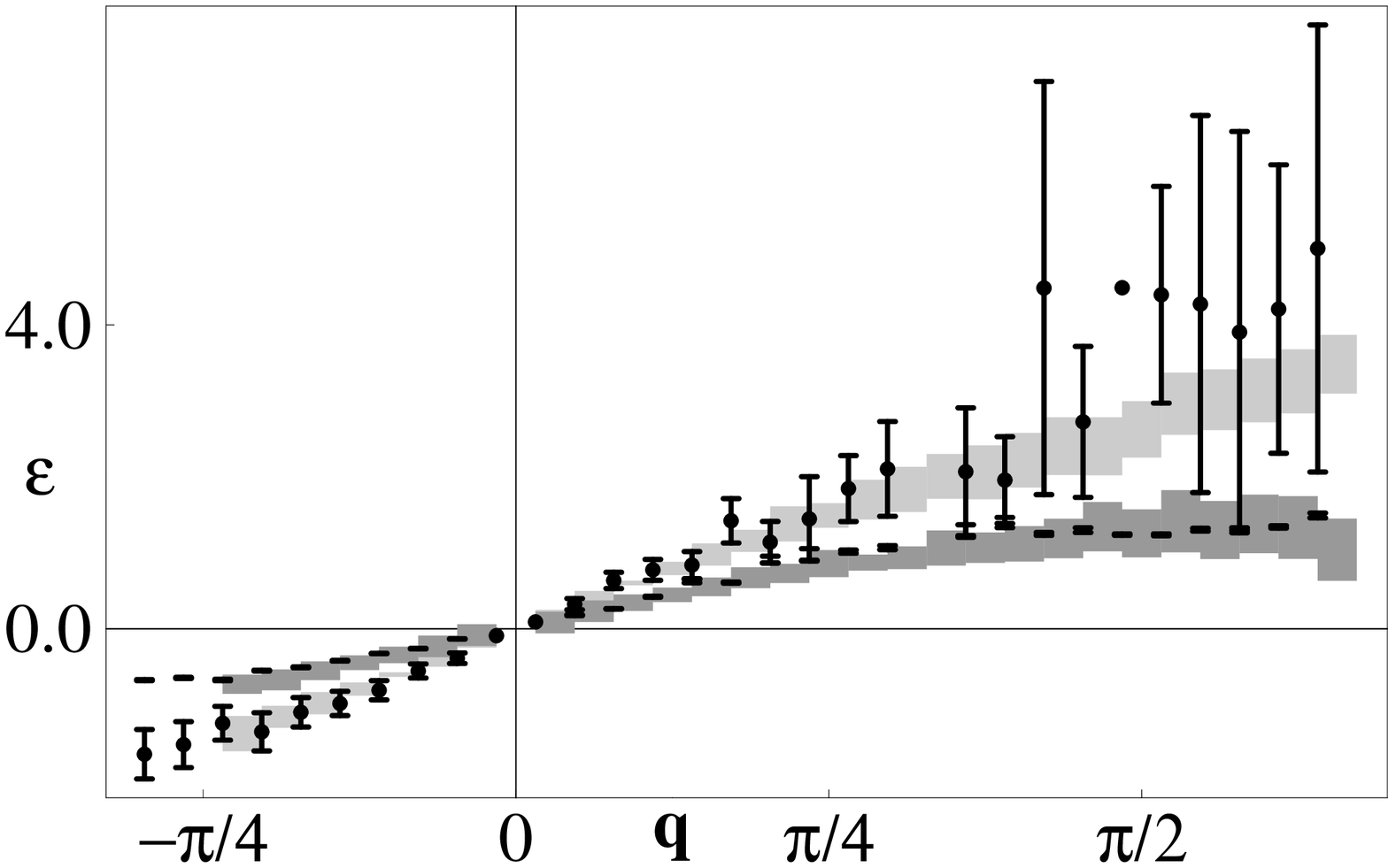}{
\caption{
Same as Fig. \figref{figfit} except that 
the fit is carried out for the data in the time interval $1.0 \le \tau \le 10.0$.
}
\label{figten}
}}

\def\figkrho{\beginfigeps{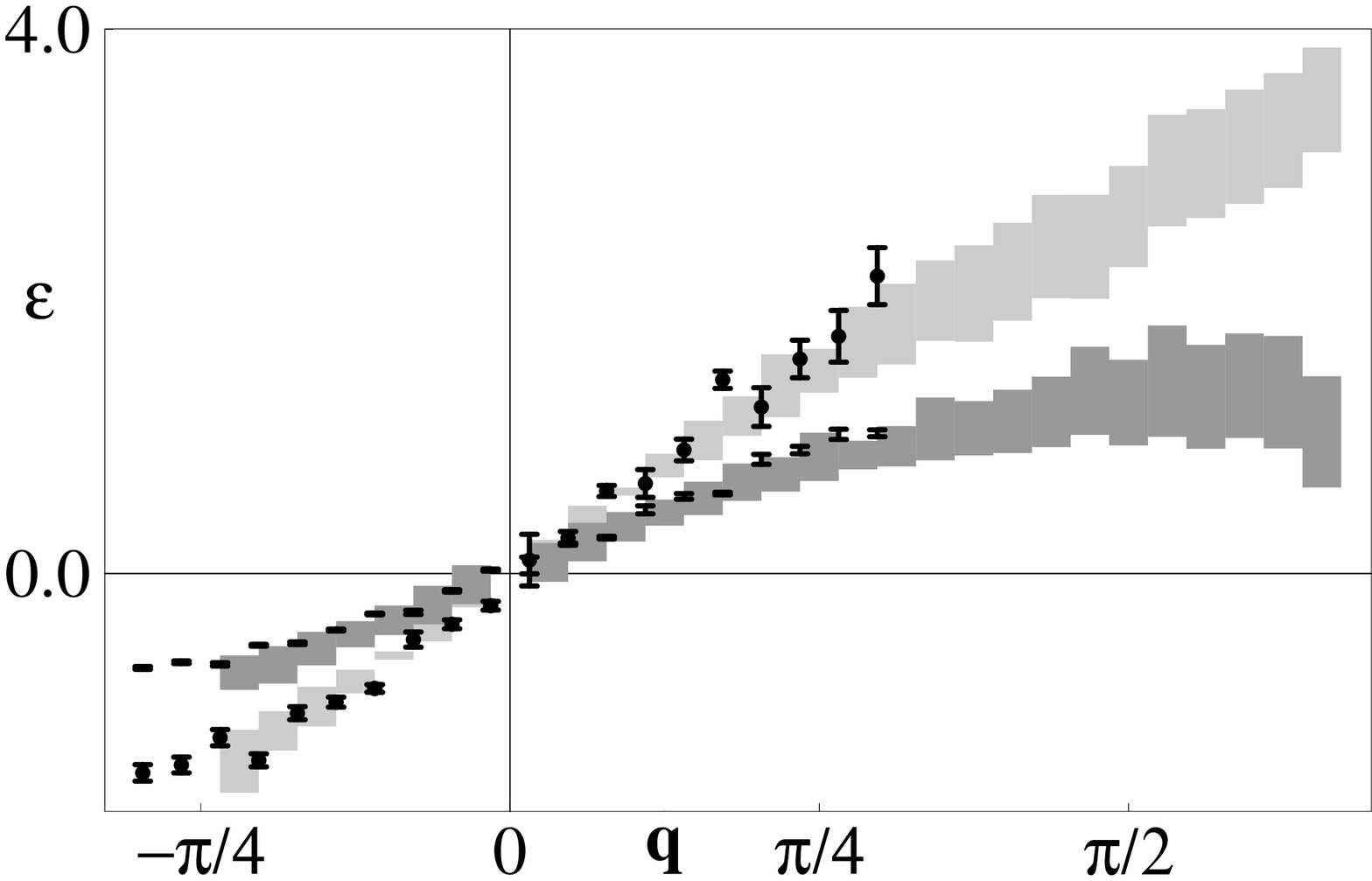}{
\caption{
Same as Fig. \figref{figfit} except that the fit is carried out with the 
 Luttinger liquid
Green's function [\eqref{glmk} with \eqref{glmx}] with 
 correlation exponent $\kr=0.7$.
}
\label{figkrho}
}}

\def\figbeta{\beginfigeps{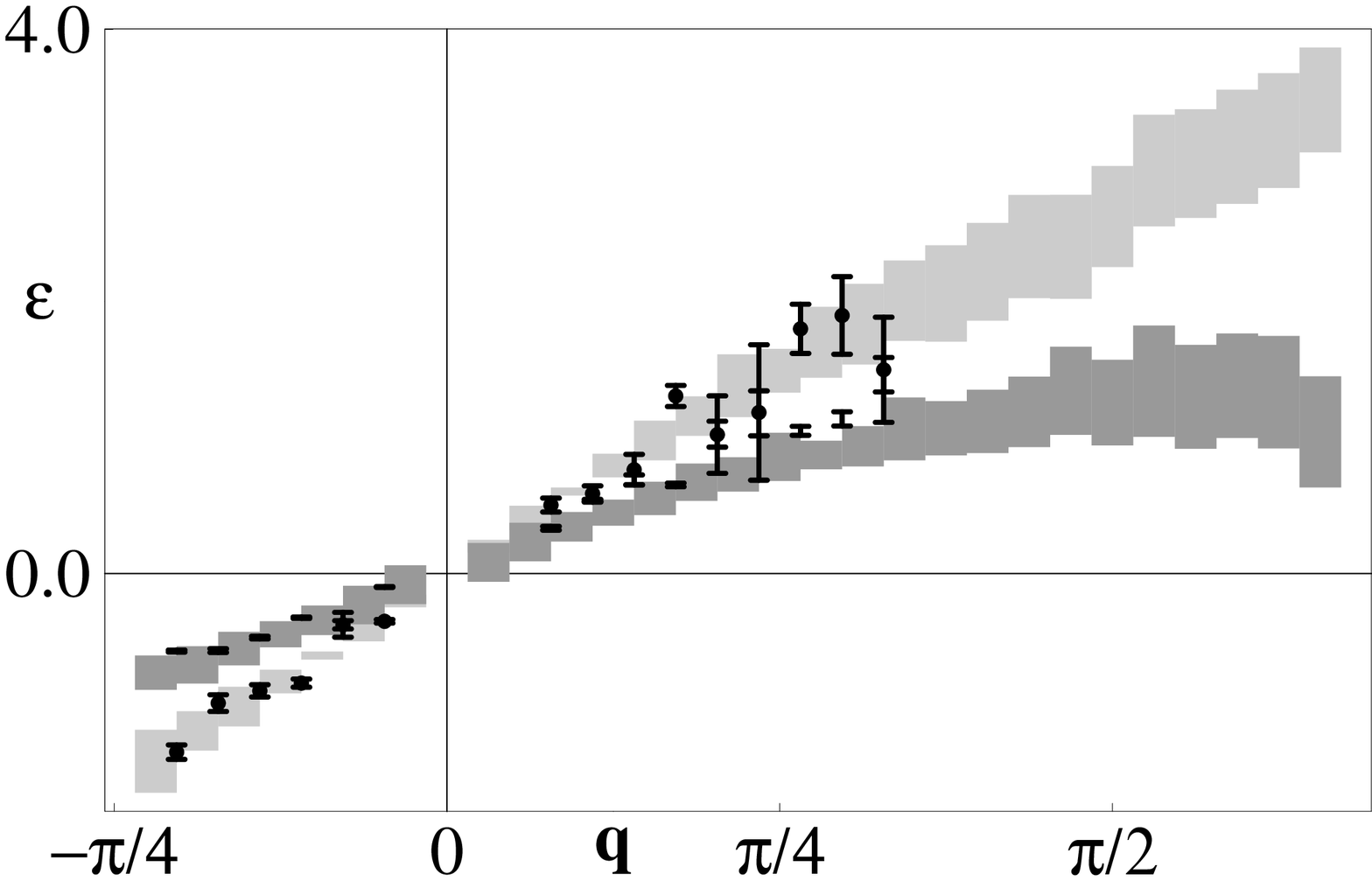}{
\caption{
Same as Fig. \figref{figfit} except that the fit is carried out with the 
 Luttinger liquid
Green's function [\eqref{glmk} with \eqref{glmxt}] with finite
 temperature ($\beta=20$)
and correlation exponent $\kr=1$.
}
\label{figbeta}
}}

\def\figgo{\beginfigeps{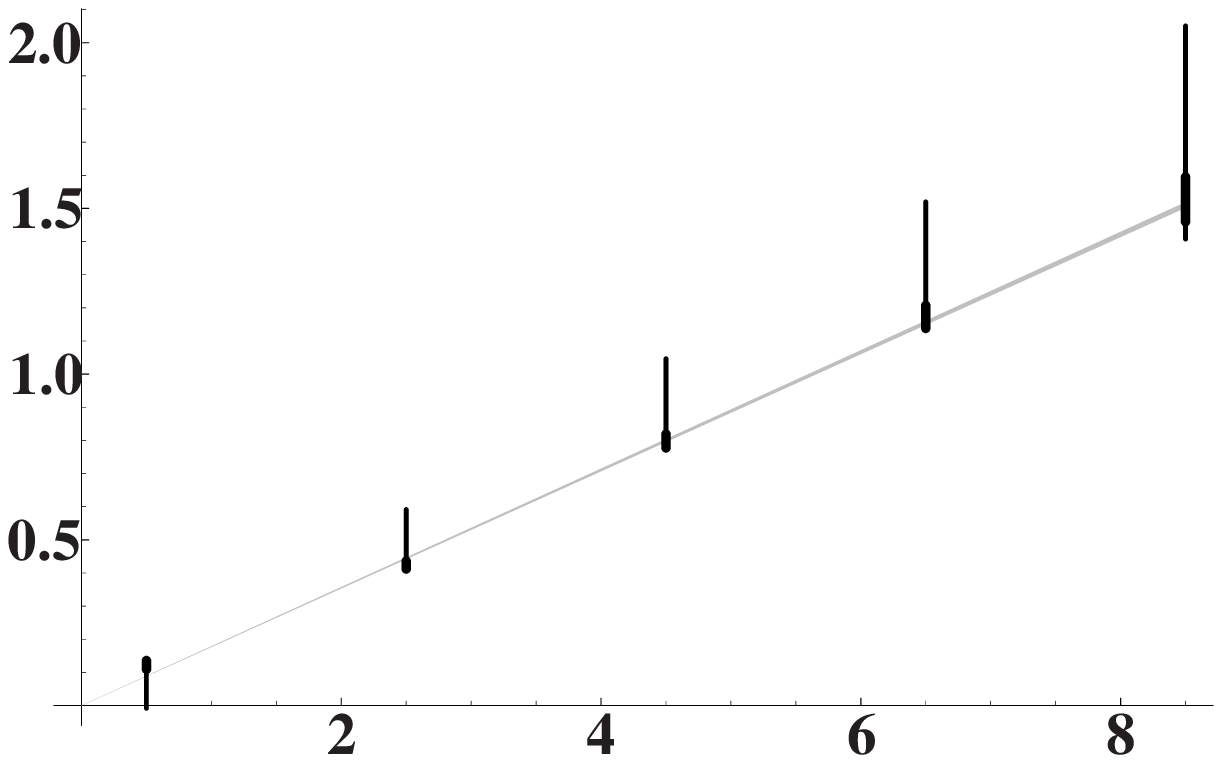}{
\caption{Fit of the $U=0$ imaginary-time Green's function with
  Eq. \eqref{glmk}.
The straight line shows $\om=v_F q$ where $v_F$ is the Fermi velocity
of the $U=0$ system.
}
\label{figgo}
}}

\def\allfigures{
\chirho
\chisigma
\ako
\aksone
\figlog
\figfit
\figten
\figkrho
\figbeta
\figgo
}

\draft

\begin{document}  

\title{ 
Systematic numerical study of spin-charge separation in one dimension
}


\author{M. G. Zacher, E. Arrigoni and  W. Hanke
}
\address{ 
Institut f\"ur Theoretische Physik,
Universit\"at W\"urzburg,
D-97074 W\"urzburg, Germany
}
\author{J. R. Schrieffer}
\address{
NHMFL and Department of Physics, Florida State University,
Tallahassee, Florida 32310}

\singlecol{

\maketitle

\begin{abstract}

The problem of spin-charge separation is analyzed numerically in the
metallic phase of the one-band Hubbard model in one dimension by
studying the behavior of the single-particle Green's function and of
the spin and charge susceptibilities.  We first analyze the
Quantum-Monte Carlo data for the imaginary-time Green's function
within the Maximum Entropy method in order to obtain the spectral
function at real frequencies.  For some values of the momentum
sufficiently away from the Fermi surface two separate peaks are found,
which can be identified as charge and spin excitations.

In order to improve our accuracy and to be able to extend our study to
a larger portion of the Brillouin zone, we also fit our data with the
imaginary-time Green's function obtained from the Luttinger-model
solution with two different velocities as fitting parameters.  The
excitation energies associated with these velocities turn out to
agree, in a broad range of momenta,
 with the ones calculated from the charge and spin susceptibilities.
This allows us to identify these single-particle excitations as due to
a separation of spin and charge.  Remarkably, the range of momenta
where spin-charge separation is seen extends well beyond the region of
linear dispersion about the Fermi surface.  We finally discuss a
possible extension of our method to detect spin-charge separation
numerically in two dimensions.

\end{abstract}


}

\section{INTRODUCTION}
\label{intro}

One-dimensional  (1D) interacting 
fermion systems show a number of anomalous
properties which cannot be understood in the framework of 
the Fermi-liquid theory
of normal metals.
In particular, their momentum distribution and density
of states  are in sharp contrast with 
 Fermi-liquid theory for energies and momenta
close to the Fermi surface.
In general,  1D systems
 can be  described by an effective low-energy theory based on 
the exactly solvable Luttinger
 model 
with suitably renormalized parameters 
and are thus referred to as
Luttinger liquids (LL)\pcite{emer.rev,hald.81,voit.rev}.
One of the most striking features of the  \LM
is the complete separation of spin and charge degrees of freedom
which manifests itself in the splitting of the single-particle spectral 
function  in two peaks corresponding to spin and charge 
excitations pro\-pagating
independently \pcite{li.wu.68,me.sc.92,voit.93,voit.rev}. 
Another important
characteristic  of the \LM is 
 the presence of power-law behavior 
with interaction-dependent exponents
for various correlation
functions. 

Beside its application to 1-D systems, 
LL theory has received 
particular 
attention in the past years in the framework of the theory of
high-T$_c$ superconductors.
The normal phase of 
 the high-T$_c$ CuO$_2$ planes 
shows in fact a number of anomalous properties 
which can be possibly understood, if one assumes that the CuO$_2$
planes are in a kind of two-dimensional LL 
state\pcite{and.prb.90,and.prl.90,and.sci.92,cl.st.95,and.97}.
In particular, it has been suggested that spin-charge separation 
could
be present also in the CuO$_2$ planes 
and that it plays an 
essential role in the way  particles 
are allowed to tunnel between
 the planes\pcite{cl.st.94}.

Numerical
methods have been proven to be crucial for the theoretical
understanding of models describing the  CuO$_2$ planes,
since
 electron correlation is rather strong in these systems and
 perturbative methods are necessarily limited.
Spin-charge separation is predicted exactly for the \LM: an {\it ideal}
exactly-solvable model.
It is thus important to test
numerically to what extent 
 spin-charge separation can occur in a one-dimensional {\it physical
   model}.
Moreover, in order to prove  the theories mentioned above, it
would be important to check whether some two-dimensional models exist,
which
display spin-charge separation.
Recently, there have been several attempts to detect spin-charge
separation in one- but also in two-dimensional models.
In 
the $U=\infty$ 1-D Hubbard model\cite{pe.ha.96}
spin-charge separation occurs in a natural way 
at {\it all} energies
(and not only at low energies like expected in a  LL) due to the exact
factorization of the wave function\pcite{og.sh.90}. 
In a numerical work Jagla et al. \cite{ja.ha.93} 
have observed the propagation in real time of
 a single-electron wave packet created at a time $t=0$ in a 1-D
 Hubbard model.
This wave packet splits up in two excitations propagating with
different velocities that can be associated with charge and spin.
In a work by W. O. Puttika and collaborators\cite{pu.gl.94}
the possibility of spin-charge separation in the 2D $t-J$ model 
has been signaled by the presence of two distinct characteristic wave
vectors for the spin and charge degrees of freedom.
Exact diagonalization of the 1-D $t-J$ model\cite{ed.oh.97} have
evidenced the presence of two peaks in the \SPSF whose positions scale with $t$ and
$J$, respectively, and have thus been identified with charge and spin
excitations. 
In another study \cite{to.ma.96}
 two peaks have been
detected in the \SPSF 
of a 1D $t-J$ model with corresponding peaks in the charge and spin
susceptibilities. 
These two peaks 
can be seen, however, only for the momenta which are immediately
next to the Fermi momentum
and thus they cannot be associated with a
dispersive spinon and holon band.
Finally, Kim and coworkers \cite{ki.ma.96}
have detected two dispersive bands for $k<k_F$ in a 1-D $t-J$ model close
to half filling.  
This is not surprising, since spin-charge separation
is in fact quite natural to expect when one hole is added beyond half
filling. The added  hole decomposes indeed in a spinless hole and
a spin misalignment which propagate with  different
velocities\pcite{ki.ma.96}. 
Nevertheless, the interesting result of that work is that the
photoemission spectrum of SrCuO$_2$, also showing two dispersive bands,
 is remarkably well reproduced
by the numerical results.

In this work, we present 
a systematic \QMC study of spin-charge separation
 {\it away} from half filling,
where Luttinger-liquid theory is expected to 
hold, in the {\it whole} Brillouin zone (BZ).
The nontrivial prediction of  
Luttinger-liquid theory is, in fact, that spin-charge separation occurs in the
{\it metallic } phase, where the band dispersion is {\it linear}.
Spin-charge separation {\it at} half filling,
as studied in the model of Ref. \onlinecite{ki.ma.96}, 
is, in our opinion, of a different nature,
since in the insulating phase the holon dispersion is {\it quadratic}
instead of {\it linear}. Of course, in the case of
Ref. \onlinecite{ki.ma.96} it was necessary to
remain in the  insulating phase, since the physical system in study was
half-filled.
%

For {\it some} values of the momentum $k$ we are able to see two peaks in the \SPSF
which correspond to the spin and charge excitations. However, due to
the limited resolution of the Maximum Entropy method , it is not
possible to resolve the two peaks in most of the BZ\pcite{pre94}.
For this reason, in the rest of the BZ we work with the imaginary-time Green's
function ${\cal G}(k,\tau)$ 
which is obtained directly from \QMC data without the need of
analytic continuation.
This has the advantage that one does not need to
introduce a further
source of error produced by  the analytic continuation to real frequencies.
 Specifically, we perform a  nonlinear $\chi^2$ fit 
of ${\cal G}(k,\tau)$  by using the solution of the \LM
  ${\cal G}^{(LM)}_{v_1,v_2,\kr}(k,\tau)$\cite{emer.rev} with two velocities
  $v_1$, $v_2$, and a normalization constant $c$ 
as fitting parameters\pcite{krho}.
Our fit yields a finite value of the difference 
$v_2-v_1$ larger than the statistical error 
 in a large portion of the
 Brillouin zone. Moreover, 
the fitted values of the corresponding excitation energies 
 $v_1 (k-k_F)$ and $v_2 (k-k_F)$ {\it coincide}, within the
statistical error, with  the spin and charge excitations, respectively, calculated
independently via the associated susceptibilities.
It is remarkable that this behavior extends well beyond the region of
linear dispersion around $k_F$ where Luttinger liquid behavior is expected.

Our paper is organized as follows. In Sec. \ref{qmc}, we introduce
the model, and we show the results of the \QMC simulation and
analytic continuation to real frequencies by means of the Maximum
Entropy method.
In Sec. \ref{fit}, we discuss and show the results of our fit of the imaginary-time
Green's function with the result from the Luttinger model.
Finally we draw our conclusions in Sec. \ref{conclusions}.

\section{QUANTUM-MONTE-CARLO SIMULATION}
\label{qmc}

We consider the 1D-Hubbard model with periodic boundary conditions
described by the following Hamiltonian:
\begin{equation}
  H = -t \sum_{i, \sigma} \left( c_{i+1,\sigma}^\dagger c_{i,\sigma}^{} + \textrm{h.c.}
\right) + U \sum_i n_{i\downarrow} n_{i\uparrow},
\end{equation}
where $c_{i,\sigma}^{(\dagger)}$ are annihilation (creation) operators for an electron at site
$i$ with spin $\sigma$ and $n_{i\sigma} = c_{i,\sigma}^\dagger c_{i,\sigma}^{}$.  The
energy scale $t$ of the model will be set to unity in the rest of the paper.

The simulations were carried out with the grand-canonical \QMC method
\cite{hir83,whi89} on a 64-site lattice with inverse temperature
$\frac1{k_B T}=\beta = 20$, Hubbard repulsion
 $U=4$
and an electron density 
 of $\langle n \rangle \approx 0.75$. 
The simulations yield the one- and two-particle Green's functions at
discrete
imaginary times 
$\tau$ with $0 \le \tau \le \beta$. We used a discretization of the
imaginary-time axis $\Delta \tau= 0.0625$
 The spectra (one-particle photoemission spectrum,
charge- and spin-susceptibilities) were then obtained by 
analytically continuing the imaginary-time results to real frequencies
by means of
the Maximum-Entropy method \cite{sil90,gub91,pre94}.

\chirho
\chisigma
Figures \ref{chirho} and \ref{chisigma} show a density plot of the charge- and
spin susceptibilities $\chi_\rho$ and $\chi_\sigma$, respectively. 
The grayscale 
gives a measure for the value of
$\chi_{\rho/\sigma} (q, \omega)$  
as a function of  momentum transfer $q$ and excitation energy
$\omega$. 
The dispersion relation for spin- and charge-excitations is defined by the maxima
of $\chi_{\rho/\sigma}$ which are 
indicated by dots with errorbars in the figure. 
A linear fit of these
maxima near $q = 0$ yields the  spin- and charge-velocities 
$v_\sigma=1.170 \pm 0.074$ and $v_\rho=2.050 \pm 0.093$
which agree very well (within the statistical error)
 with Bethe-Ansatz  results \cite{sch91} for the infinite
lattice and zero temperature.

However, it is not sufficient to have two different velocities (or,
equivalently, energy
dispersions) for the two-particle spin and charge modes in order to
conclude that the system shows spin-charge separation. In fact, in a Fermi
liquid there are spin and charge excitations that 
 originate from collective modes and do not destroy the quasiparticle \cite{pines.nozie}.
The quasiparticles 
 thus remain 
well-defined and do not split into a charge and a spin excitation as it
occurs in a Luttinger liquid.
On the other hand, in a Luttinger liquid (or in  spin-charge
separated system in general)
 a particle injected at a certain point $x$
decays into a spinon and a holon  propagating
with different velocities. The separation of the two excitations could then be 
detected by means of a ``diagnostic operator'' measuring the time
dependence of spin and charge  at a given point $y$
far away from $x$. In the case of spin-charge separation, this
diagnostic operator would then
 measure two different passing times for
the
charge and spin  perturbations of the injected particle. 
\ako
True spin-charge separation in the sense of the \LM  should be thus
identified with
 different energy dispersions in the spin and charge susceptibilities
 {\it associated with} corresponding low-lying excitations in the
 single-particle 
spectrum \pcite{to.ma.96}.

In Fig.~\ref{ako} we plot this single-particle spectrum $A(k,\omega)$\cite{pre94} 
in the whole
Brillouin zone. 
Close to the Fermi momentum the band dispersion is approximately
linear, which justifies  
the mapping to the Luttinger model.
However, 
the spectrum becomes broader when going away from the Fermi surface. 
This phenomenon has two reasons: 
First, the resolution of the maximum-entropy
method gets worse at higher energies,
due to the exponential
kernel in the spectral theorem, 
and second, according to Luttinger liquid theory,
 the single peak starts to 
  split into two
peaks representing the spin- and charge-excitations propagating with different
velocities. 
However, for $k$ very close to $k_F$ 
these two peaks, which should be separated by an energy
$(v_2-v_1)(k-k_F)$,
are still too close together for the 
 Maximum Entropy method to distinguish them.
On the other hand, at larger values of $(k-k_F)$
the excitation energies are too high and the maximum entropy method
becomes less reliable as explained above.
In both these cases the two peaks 
merge into a single broader peak and spin-charge separation is not detectable.
There are, however, some favorable intermediate $k$-points 
where
spin-charge separation is directly detectable in the single-particle
spectral function. 
In  figure \ref{aks1} we show the spectral function for one of these
favorable points.
Here $k$ is neither too close nor too far from the Fermi surface and the
maximum-entropy method (without using any {\it prior knowledge})
yields two well-separated peaks. Their positions are consistent with the
spin and charge excitation energies 
 (indicated by two dots
with horizontal errorbars)
calculated independently
 from the spin- and charge-
velocities ($\omega_{\rho/\sigma} = \Delta k \, v_{\rho/\sigma}$).
Previously, it was not possible to resolve
spin-charge separation in the one-particle spectrum\cite{pre94}   mainly because they were
carried out in a low doping regime ($\langle n \rangle$ close to 1) where the difference of
spin- and charge-velocities is relatively small \pcite{sch91}.
\aksone

\section{FIT OF THE IMAGINARY-TIME GREEN'S FUNCTION}
\label{fit}

In order to carry out a {\it systematic} study of spin-charge
separation it is important to detect spin and charge excitations over the
whole BZ, or at least in an extended region around the Fermi surface.
However, due to the additional  rather large error introduced by the
Maximum Entropy analytic continuation method
 to the \QMC data, this
turns out to be very difficult for many $k$ points, as we have
discussed above.
For this reason, we
work directly with the data for the {\it imaginary-time} Green's function
${\cal G}(k,\tau)$. 
In the asymptotic limit ($\tau \gtrsim 1$) and close to the Fermi
surface ($+ k_F$)
this function should approach the 
 Green's function 
of the \LM for  right-moving fermions, i. e. 
\beq
{\cal G}^{(LM)}_{v_1,v_2,\kr}(k,\tau) \equiv \int dx \ e^{-i k x} 
\tilde{\cal G}^{(LM)}_{v_1,v_2,\kr}(x,\tau) \;, 
\label{glmk}
\eeq
with 
\beqn
\label{glmx}
 && \tilde{\cal G}^{(LM)}_{v_1,v_2,\kr}(x,\tau)  
\\ \nonumber
&& = \frac {e^{i k_F x} c}{\sqrt{v_1 \tau + i x}\sqrt{v_2 \tau + i x}(x^2+v_2^2
  \tau^2)^{-(\kr+1/\kr-2)/8}} \;,
\eeqn
where $c$ is a normalization constant\cite{norm} and $k_F$ the Fermi momentum.
 Therefore, 
in order to identify the spin and charge excitations {\it directly} in
the Green's function,
we 
carry out a nonlinear $\chi^2$
fit of our data for ${\cal G}(k,\tau)$
to ${\cal G}^{(LM)}_{v_1,v_2,\kr}(k,\tau)$.
The fit parameters are the
two velocities $v_1$ and $v_2$, 
and the normalization constant $c$ \pcite{krho}.
Due to the statistical error in the \QMC data, we get  
statistical errors 
$\Delta v_1$, $\Delta v_2$, and $\Delta c$
for the parameters 
$v_1$, $v_2$, and $c$, respectively.
The splitting of the single-particle mode into
two excitations is
 thus detected when the difference between
the two velocities is larger than the statistical error.
Furthermore, in order to make sure that the two excitations coincide with
the spin and the charge modes one has to compare $v_1$ and $v_2$ with
the velocities $v_{\rho}$ and $v_{\sigma}$ calculated independently
via the susceptibilities.

\figlog
However, in order to carry out this fit
one should not use the data from the whole interval 
$0.0 \le
\tau\le \beta$ for the following reasons. First of all the 
Hubbard model and the Luttinger model Green's function should coincide
only asymptotically. For this reason, we choose $\tau \ge
1.0$. Moreover, the interval $\beta/2 \le \tau\le \beta$ is equivalent
to the one $- \beta/2 \le \tau\le 0$ so that we can omit the former.
In addition, as can be seen in 
 Fig. \ref{figlog} the log of the imaginary-time Green's 
 is
quite sharply defined up to $\tau\approx 5.0$. For $\tau\cmag 5.0$,  large (relative) errors
start to develop due to the small value of the Green's function in these
points.
For this reason, we choose to carry out our fit {\it} only for
the data in the interval $1.0 \le \tau\le 5.0$ in order to select the
less ``noisy'' data.
In order to check that our results do not depend on this choice, we
also carry out the 
fit for the data in the interval $1.0 \le
\tau\le \beta/2=10.0$ (Fig. \ref{figten}). This turns out to be quite similar to
the first one (Fig. \ref{figfit})
except for larger statistical errors.
In Fig. \ref{figfit} we show the result of our fit with the $T=0$
Green's function [\eqref{glmk} with \eqref{glmx}] for several values
of $q=(k-k_F)$. The vertical black lines show the value of the
spinon and holon excitation energies $\eps_1(q)\equiv v_1(q) \ q$ and
$\eps_2(q) \equiv v_2(q) \ q$,
respectively, obtained from the fit with the single-particle Green's
function\pcite{evsv}. As one can see, we obtain a clear separation of the two
modes for almost all the $q$ points.
In addition, the velocities are slightly $q$-dependent as expected from a curved band.
 To check that these modes correspond
to spin and charge degrees of freedom, we plot in the same figure   the
dispersions 
 calculated from the peaks of the
 corresponding susceptibilities. The width of the gray regions
 indicate the peak positions within their uncertainty.
 As one
can see, the dispersions obtained in the two ways coincide within the statistical error.
We find it remarkable that, even at $k$-points far from $k_F$,
the fit with the  Luttinger-liquid  Green's
function 
 is in agreement with the two-particle
response, although the dispersion is no longer linear. 
It thus seems that spin and charge separation survives
even at  higher energies. 

\figfit
\figten
In Fig. \ref{figfit} we used the simplest form of $\eqref{glmx}$,
namely, the one with $\kr=1.0$.
This  is not,
in principle, the correct value of the correlation exponent
$\kr$ when $U \not=0$.
Actually, one could
use $\kr$ as a
further parameter to fit the data
or, alternatively, use the 
result from the Bethe-Ansatz solution\cite{schu.91}. 
It turns out,
 however, that an attempt to fit $\kr$  yields an 
 error  of the order of $0.5$, which means that $\kr$ cannot
 be determined by our fit. 
It also turns out that 
the {\it result} of
 the fit does not depend crucially on the value of $\kr$ we are
 using. Indeed, as one can see  
from Fig. \ref{figkrho}, where we show the results of
the fit 
obtained with the Bethe-Ansatz value
 $\kr=0.7$
the results of the fit
 do not differ appreciably from the ones in Fig. \ref{figfit}.
\figkrho
For this reason, the non-interacting value $\kr=1.0$ can be safely used.
This is important, because in this way
 it is possible to
test the occurrence of
 spin-charge separation  even {\it without} knowing whether the system
 has anomalous scaling ($\kr \not=1$) 
or not.
 This could be useful 
in  cases where
 the anomalous exponent may be not known a priori and  may be
 difficult to evaluate. 
In this case, one can  assume a form of the fitting
 function without spectral anomaly (i. e., $\kr=1$), but simply with a branch cut due to
 spin-charge separation. This could be useful, for example, to test
 spin-charge separation in 2-D as we shall discuss below.

Since the \QMC simulations are carried out at finite temperature 
$\beta=1/(k_B T)=20$, we also perform our fit with the 
\LM Green's function at the same temperature 
${\cal G}^{(LM)}_{v_1,v_2,\kr}(k,\tau ; \beta=20)$, 
which has the same form as \eqref{glmk} with \eqref{glmx} replaced
with
\beqn
&& \tilde{\cal G}^{(LM)}_{v_1,v_2,\kr}(x,\tau; \beta)  
\\ \nonumber &&
 =  e^{i k_F x}
 \sqrt{g_{v_1,1}\left(\frac{\pi x}{\beta},\frac{\pi \tau}{\beta} ;
     \beta \right) \;
   g_{v_2,\kr}\left(\frac{\pi x}{\beta},\frac{\pi \tau}{\beta} ; \beta\right)} \;,
\label{glmxt}
\eeqn
with
\beqn
&& g_{v,K}(\tilde x,\tilde \tau ; \beta)  =
\\ \nonumber &&
-i \left[\frac{\beta}{\pi}\left(
 \cosh(\tilde x)^2 - \cos(\tilde \tau)^2\right)\right]^{-\frac{K+1/K}{4}}
e^{-i \arg(\tanh \tilde x + i \tan \tilde \tau)} \;.
\eeqn
In Fig. \ref{figbeta}, we show the fit performed with the more
complicated 
finite-temperature ($\beta=20$)  Green's function  \eqref{glmxt}. As
one can see, the results are not appreciably different from the ones
of Fig. \ref{figfit}, which means that the temperature of our
simulation is low enough and we can safely fit our results with the
$T=0$ Green's function. 
\figbeta

Finally, to check that the two different velocities $v_1$ and $v_2$
obtained
are not an artifact of our fit, we carry out a fit of 
the {\it
  non-interacting} Green's function $\tilde{\cal
  G}^{(0)}(k,\tau;\beta)$ with the function \eqref{glmk}
assuming artificially the same statistical
errors as the ones obtained in the \QMC simulation. As one can see in
Fig. \ref{figgo},
in this case 
the spin and charge velocities obtained are equal within the statistical error, as
it should be.
\figgo

Another 
motivation of this work was to test a ``diagnostic
operator'' that can be applied to detect numerically the occurrence of 
spin and charge separation in a many-body system from \QMC data.
If one uses exact diagonalization,  
where 
the spectral function of the system can be evaluated directly, 
 it is of course
not necessary to fit the imaginary-time Green's
function. However, we believe that the systems that can be studied by
exact diagonalization ($10$-$16$ sites) are too small to allow for a systematic study of
spin-charge separation (except for a very high value of the momentum,
like in Ref.\onlinecite{ja.ha.93}).
The Fourier transform 
(in momentum and imaginary-frequency space)
of the 
spin-charge separated Green's function Eq. \eqref{glmx} with
$\kr=1$ reads 
\beq
\hat{\cal G}^{(LM)}_{v_1,v_2,\kr}(k,\om) \propto 
\frac 1{\sqrt{i \om - \eps_1(k)}\sqrt{i \om - \eps_2(k)}} \;,
\label{glmxw}
\eeq
where $\eps_1(k)=v_1 (k-k_F)$ and
 $\eps_2(k)=v_2 (k-k_F)$ represent the spin and charge excitations
(measured from the chemical potential) 
in which the single-particle excitation is split.
The same form of $\hat{\cal G}^{(LM)}_{v_1,v_2,\kr}(k,\om)$ could be
expected to hold asymptotically, i.e. for small frequencies and close
to the Fermi surface, in higher dimensions. Close to the Fermi surface, one will
have a direction-dependent dispersion 
$\eps_i(\vec k)= (\vec k - \vec k_F) \cdot \vec v(\vec k_F)$
where $v(\vec k_F)$ is the Fermi velocity of the point at the Fermi
surface $\vec k_F$ closest to $\vec k$.
Spin-charge separation
would be signaled by two different, direction-dependent $\eps_1(\vec k)$
and $\eps_2(\vec k)$
for a given $\vec k$.

\section{CONCLUSIONS}
\label{conclusions}

To summarize, we carried out a test of spin-charge
separation in the 1-D Hubbard model at finite doping. 
 It is in
general difficult to resolve the peaks corresponding to the spin and
charge
 excitations in the
single-particle spectral function due to the loss of accuracy which
occurs when analytically continuing the imaginary-time \QMC results to
real frequencies.
For some values of the momentum close to the Fermi surface, however, 
 we were able to resolve two  peaks  whose energies correspond to the
peaks at the same $q=k-k_F$ in the spin and charge susceptibilities, respectively.

By fitting the
\QMC data for the imaginary-time Green's function with the exact
solution from the Luttinger model with the spin and charge
velocities as fitting parameters, we have been able to
resolve the two excitations over the whole Brillouin zone.
The two excitation energies found in the fit agree, within
statistical error, with the spin
and charge excitations, respectively, identified with the peaks 
of the spin and charge susceptibilities.
Remarkably, this occurs
 also away
from the region where the band dispersion is linear.
We also suggested a possible extension of this ``diagnostic operator'' to
test a possible occurrence of spin-charge separation in two dimensions.

\section{Acknowledgments}

We thank
 J. Voit, W. von der Linden, F. Mila, and M. Imada for stimulating
discussions
and precious suggestions.
This work was partially supported  by the Bavarian high-Tc program FORSUPRA 
and by the 
BMBF (05 605 WWA 6).
M.\ G.\ Z.\ acknowledges support from the DFN project Tk598-VA/D3. 
E.\ A.\ 
 acknowledges research support from the TMR
  program  ERBFMBICT950048 of the European Community.
J.\ R.\ S.\ acknowledges support from NSF grant DMR-9629987.
The numerical calculations were carried out at the HLRZ J\"ulich.

\ifx\undefined\andword \def\andword{and} \fi 
\ifx\undefined\submitted \def\submitted{submitted} \fi 
\ifx\undefined\inpress \def\inpress{in press} \fi 
\def\nonformale#1{#1}
\def\formale#1{}
\def\spa{} \def\spb{}
\def\spa{\ifpreprintsty\else\vspace*{-.5cm}\fi} 
\def\spb{\ifpreprintsty\else\vspace*{-1.6cm}\fi} 
\spa

\showfigures


\begin{thebibliography}{10}
\bibitem{norm}{
It is not possible to fix the constant $c$ through sum rules, since the two
  Green's functions coincide only asymptotically.}
\bibitem{evsv}{
The $q$-dependence of the velocities is due to the fact that we fit
  independently ${\cal G}(k_F+q,\tau)$ for each $q$.}
\end{thebibliography}
\end{document}